\documentclass[superscriptaddress,showpacs,amssymb,10pt,reprint,aps,prd,longbibliography,nofootinbib,floatfix]{revtex4-1}

\usepackage{graphicx,epsfig,amssymb} 
\usepackage{amsmath,amsfonts, times}
\usepackage{bm} 

\usepackage{epstopdf}
\usepackage[linktocpage,colorlinks]{hyperref}
\usepackage[caption=false]{subfig}
\usepackage[usenames]{color}     
\usepackage{natbib}
\usepackage{soul}
\usepackage[utf8x]{inputenc}
\usepackage[usenames]{color}
\definecolor{coolblack}{rgb}{0.0, 0.18, 0.39}
\definecolor{darkred}{rgb}{0.5,0,0}
\definecolor{darkgreen}{rgb}{0,0.5,0}
\definecolor{darkblue}{rgb}{0,0,0.5}
\definecolor{lapislazuli}{rgb}{0.15, 0.38, 0.61}
\definecolor{venetianred}{rgb}{0.78, 0.03, 0.08}
\definecolor{bleudefrance}{rgb}{0.19, 0.55, 0.91}
\definecolor{dogwoodrose}{rgb}{0.84, 0.09, 0.41}
\hypersetup{colorlinks=true, citecolor=darkgreen, linkcolor=darkblue, 
	urlcolor = blue}

\def\p{\partial}

\begin{document}
	\title{\large Shadows of charged rotating black holes: Kerr-Newman versus Kerr-Sen}

	\author{S\'ergio V. M. C. B. Xavier}
	\email{sergio.xavier@icen.ufpa.br}
	\affiliation{Programa de P\'os-Gradua\c{c}\~{a}o em F\'{\i}sica, Universidade 
		Federal do Par\'a, 66075-110, Bel\'em, Par\'a, Brazil.}
	
	\author{Pedro V. P. Cunha}
	\email{pedro.cunha@aei.mpg.de}
	\affiliation{Max Planck for Gravitational Physics - Albert Einstein Institute,\\
		Am Mühlenberg 1, Potsdam 14476, Germany\\
	}

	\author{Lu\'is C. B. Crispino}
	\email{crispino@ufpa.br}
	\affiliation{Programa de P\'os-Gradua\c{c}\~{a}o em F\'{\i}sica, Universidade 
		Federal do Par\'a, 66075-110, Bel\'em, Par\'a, Brazil.}
	
	\author{Carlos A. R. Herdeiro}
	\email{herdeiro@ua.pt}
	\affiliation{Departamento de Matem\'atica da Universidade de Aveiro\\ and Centre for Research and Development in\\ Mathematics and Applications (CIDMA),\\ Campus de Santiago, 3810-183 Aveiro, Portugal\\
	}

	\begin{abstract}
		Celebrating the centennial of its first experimental test, the theory of General Relativity (GR) has successfully and consistently passed all subsequent tests with flying colours. It is expected, however, that at certain scales new physics, in particular in the form of quantum corrections, will emerge, changing some of the predictions of GR, which is a classical theory. In this respect, black holes (BHs) are natural configurations to explore the quantum effects on strong gravitational fields. BH solutions in the low-energy effective field theory description of the heterotic string theory, which is one of the leading candidates to describe quantum gravity, have been the focus of many studies in the last three decades. The recent interest in strong gravitational lensing by BHs, in the wake of the Event Horizon Telescope observations, suggests comparing the BH lensing in both GR and heterotic string theory, in order to assess the phenomenological differences between these models. In this work, we investigate the differences in the shadows of two charged BH solutions with rotation: one arising in the context of GR, namely the Kerr-Newman solution, and the other within the context of low-energy heterotic string theory, the Kerr-Sen solution. We show and interpret, in particular, that the stringy BH always has a larger shadow, for the same physical parameters and observation conditions.
	\end{abstract}
	
	\date{\today}
	
	\maketitle

	\section{Introduction}\label{sec:int}
	\hspace{0.5cm} Recently, the international Event Horizon Telescope (EHT) collaboration has unveiled the first shadow image of a supermassive black hole (BH), located in the center of the Messier 87 galaxy \cite{M87_1, M87_2, M87_3, M87_4, M87_5, M87_6}. In general, the BH shadow is determined by a special set of bound photon orbits, which are a manifestation of the BH's strong bending of light rays around the BH \cite{Cunha:2017eoe}. A class of these orbiting trajectories exist for the Kerr spacetime, being dubbed spherical photon orbits (SPOs)~\cite{Teo}. Although such orbits are unstable outside the horizon, they are important because they define the capture threshold of light rays during a scattering process around a BH. The photons escaping from these spherical orbits towards an observer determine the edge of the BH shadow as seen by this observer \cite{Cunha_Herdeiro:2018}. Together with the LIGO-Virgo detections of gravitational waves \cite{LIGO_1,LIGO_2,LIGO_3,LIGO_4,LIGO_5} and the examinations of x-ray binaries \cite{Shakura:1973, Tetarenko:2016, Barack_etal:2019}, the EHT observations became a tool to test strong gravitational fields.

	Since $1919$ \cite{Dyson_etal:1920,Crispino_Kennefick:2019} the phenomenon of gravitational lensing has been investigated in General Relativity (GR) theory. In particular, the shadow of rotating BHs attracted a lot of attention in the last decades, beginning with the analysis of the Kerr shadow \cite{Bardeen:1973} following Kerr-like metrics without and with optically thin accretion disks \cite{Luminet:1979,Falcke_etal:2000, Johannsen_Psaltis:2010}, and some other configurations of rotating BHs \cite{Cunha_etal:2015,Cunha_etal:2016, Abdujabbarov_etal:2016, Cunha_etal:2017, Amarilla_Eiroa:2000, Abdujabbarov_etal:2013, Amarilla_Eiroa:2013, Papnoi:2014, Amir_Ghosh:2016, Dastan_etal:2016, Younsi:2016azx, Eiroa_Sendra:2018,Schee:2009,Schee:2018,Kumar:2018ple,Gibbons_etal:2016,Cvetic:2017zde}. One of the motivations of the several works on this subject is the fact that trajectories of light near BHs and other compact objects are related to important features and properties of the background geometry. For instance, the quasinormal modes of BHs have been associated with the parameters of the unstable photon geodesics \cite{Cardoso_etal:2009, Hod:2009, Dolan:2010,Konoplya:2017wot}. Moreover, the spin, the mass parameter and possible other global charges or ``hair" parameters of the BH can, in principle, be inferred by the observation of the size and deformation of its shadow  \cite{maeda, Tsupko:2017,Cunha:2019dwb,Cunha:2019ikd}. 
	
	Besides the fact that GR has passed numerous experimental tests, it is expected that at sufficiently small scales GR will breakdown,  in particular when quantum effects are expected to become relevant. There are several motivations to modify GR, from the formation of physical singularities inside BHs, to non-renormalizability of the theory. Among the attempts to quantize gravity, string theory holds as one of the most promising candidates~\cite{Polchinski}, with most of the analyzes therein having focused on the low-energy limit. 
	
	One of the most interesting scenarios to explore the quantum nature of strong gravitational fields is a BH environment \cite{Callan_etal:1988}. In $1992$, in the context of the low-energy limit of heterotic string theory, a rotating BH solution has been found by Sen, which became known as the Kerr-Sen BH (KSBH) \cite{Sen1992}. In addition to the mass $M$ and angular momentum $J$, the KSBH  has a third physical parameter: the electric charge $Q$. Recently, various aspects of KSBH physics have been discussed, for instance radial geodesics \cite{Blaga:2001}; the capturing and scattering of photons \cite{Hioki:2008}; hidden conformal symmetries \cite{Ghezelbash:2013}, the shadow of KSBH for an observer at infinity, both in vacuum and in the presence of a plasma \cite{Dastan_etal:2016}. Furthermore, the  cosmic censorship conjecture \cite{Gwak:2017} and Hawking temperature \cite{Khani:2013, Khani:2014} have also been investigated for the Kerr-Sen spacetime, as well as many other aspects, see $e.g.$,  Ref.~\cite{Cayuso:2019ieu}.
	
	In order to understand possible imprints of string theory in an astrophysical environment, even if as a toy model, as electric charge is expected to be negligible for astrophysical BHs, we can search for observational signatures present in the image of the shadow of a BH. Therefore, it is instructive to obtain the shadow of KSBHs and compare the results with those of the Kerr-Newman BHs (KNBHs), a charged and rotating solution in the Einstein-Maxwell theory. One can quantify this difference by calculating the radius of the shadows in each case. This comparison is performed in this paper. We show, in particular, that the stringy BH always has a larger shadow, for the same physical parameters and observation conditions and offer a possible interpretation for this result.  Throughout this paper, unless otherwise stated, we adopt natural units $(c=G=\hbar=1)$.

	\section{Kerr-Newman black hole} \label{sec:1}
	The Einstein-Maxwell theory, in eletrovacuum GR, is described by the following action:
	\begin{equation}
	S= \frac{1}{4\pi}\int d^4x\sqrt{-g}\left(\frac{R}{4}-\frac{1}{4}F_{\mu\nu}F^{\mu\nu}\right), 
	\end{equation}
	where $R$ is the Ricci scalar curvature, $g$ is the determinant of the metric with covariant components $g_{\mu\nu}$ and $F_{\mu\nu}$ is the electromagnetic tensor obtained from
	\begin{equation}
	F_{\mu\nu}=\partial_{\mu}A_{\nu}-\partial_{\nu}A_{\mu},
	\label{Fmunu}
	\end{equation}
	with $A_{\mu}$ being the electromagnetic vector potential. The Kerr-Newman (KN) spacetime is the most general stationary asymptotically flat BH solution in four dimensional spacetime in the Einstein-Maxwell theory \cite{Mazur:1982}.

	Obtained in $1965$ by Newman and collaborators, the KN metric is a charged generalization of the Kerr solution \cite{Kerr:1963}. This solution of the Einstein-Maxwell equations describes a stationary electrically charged rotating BH and was obtained after applying a complex transformation to the Reissner-Nordstr\"om (RN) solution \cite{Newman&Jannis}.
	
	Written in Boyer–Lindquist coordinates $\{t, r, \theta, \phi\}$, the line element of the KNBH is given by:
	\begin{align}
	ds^{2}=&-\left(1-\frac{2 M r - Q^{2}}{\rho^{2}_{KN}}\right)dt^{2}+\frac{\rho^{2}_{KN}}{\Delta_{KN}}dr^{2}+\rho^{2}_{KN}d\theta^{2}-\nonumber\\
	&\frac{4 Mar\sin^{2}\theta-2aQ^{2}\sin^{2}\theta}{\rho^{2}_{KN}}dtd\phi+\nonumber\\
	&\frac{\left[(r^{2}+a^{2})^{2}-\Delta_{KN} a^{2}\sin^{2}\theta\right]\sin^{2}\theta}{\rho^{2}_{KN}}d\phi^{2},
	\end{align}	
	and the electromagnetic vector potential is
	\begin{equation}
	A_{\mu}dx^{\mu}=-\frac{Qr}{\rho^{2}_{KN}}(dt-a\sin^{2}\theta d\phi), 
	\label{A_KN}
	\end{equation}		
	where
	\begin{align}
	&\Delta_{KN}\equiv r^{2}-2 M r+a^{2}+Q^{2},\\
	&\rho^{2}_{KN}\equiv r^{2}+a^{2}\cos^{2}\theta.
	\end{align}
	$M$, $a$ and $Q$ are the mass, the angular momentum per unit mass and the electrical charge of the BH, respectively. The event horizon of the KNBH is located at 
	\begin{equation}
	r^{KN}_{+}=M+\sqrt{M^{2}-Q^2-a^2}.
	\label{rH_KN}
	\end{equation}
	In the limit of $a\rightarrow 0$ the KN metric reduces to the Reissner-Nordstr\"om solution, whereas the limit $Q\rightarrow 0$ yields the Kerr metric. For more details of both limits see Ref.~\cite{chandrasekhar}. To ensure the existence of the event horizon, we shall restrict our analysis to the case $a^{2}+Q^{2}\leqslant M^{2}$. The case $a^{2}+Q^{2}> M^{2}$ contains a naked singularity.

	
	In 1968 Brandon Carter found that in the KN spacetime there exists a separation constant~\cite{Carterconstant} such that the Hamilton--Jacobi equation is separable for geodesics and the geodesic motion is Liouville integrable. The existence of this additional constant of the motion, which is quadratic in the momentum, is not expected from the spacetime isometries. It is in one to one correspondence with the existence of a non-trivial (irreducible) rank 2 Killing tensor in the Kerr geometry. The latter, in turn, is implied by the existence of a more fundamental object called a Killing-Yano tensor. If the Killing-Yano tensor is non-degenerate, its existence implies that the spacetime is of Petrov type D \cite{Frolov:2017kze}. The equations of motion for a null geodesic propagating in the KN spacetime are given by:
	\begin{align}
	&\rho^{2}_{KN}\dot{r}=\pm\sqrt{R_{KN}},
	\label{KNrcarter}\\
	&\rho^{2}_{KN}\dot{\theta}=\pm\sqrt{\Theta_{KN}},
	\label{KNthetacarter}\\
	&\rho^{2}_{KN}\dot{t}=\frac{\left(r^{2}+a^{2}\right)^{2}-a \Phi_{KN}(r^{2}+a^{2})}{\Delta_{KN}}-\nonumber\\
	&\hspace{1cm}\sin^{2}\theta\left(a^{2}-\frac{a\Phi_{KN}}{\sin^{2}\theta}\right),
	\label{KNtcarter}\\
	&\rho^{2}_{KN}\dot{\phi}=\frac{a(r^{2}+a^{2})-a^{2}\Phi_{KN}}{\Delta_{KN}}-\frac{a\sin^{2}\theta-\Phi_{KN}}{\sin^{2}\theta},
	\label{KNphicarter}
	\end{align}
	where 
	\small
	\begin{align}
	&R_{KN}\equiv\left[(r^{2}+a^{2})-a\Phi_{KN}\right]^{2}-\Delta_{KN}\left[(\Phi_{KN}-a)^{2}+\eta_{KN}\right],\\
	&\Theta_{KN}\equiv\left[\eta_{KN}+(\Phi_{KN}-a)^{2}\right]-\frac{(a\sin^{2}\theta-\Phi_{KN})^{2}}{\sin^{2}\theta}.
	\end{align}
	\normalsize
	The impact parameters $\Phi_{KN}\equiv L/E$ and $\eta_{KN}\equiv K/E^{2}$ are defined using the photon's azimuthal angular momentum $L$, the Carter's constant $K$ and energy $E$, which are geodesic constants of motion. 
	
	For the case of SPOs~\cite{Teo}, which are bound photon orbits with a constant Boyer-Lindquist radial coordinate, and which determine the edge of the BH shadow,  these parameters can be expressed as functions of the radial coordinate of the corresponding SPO as follows:
	\begin{align}
	&\Phi_{KN}=\frac{a^2 M+a^2 r-3 M r^2+2 Q^2 r+r^3}{a (M-r)},\\
	&\eta_{KN}=-\frac{r^2 \left[4 a^2 \left(Q^2-M r\right)+\left(r (r-3 M)+2 Q^2\right)^2\right]}{a^2 (M-r)^2}.
	\end{align}

	For stationary and axi-symmetric BH spacetimes, one may unambiguously define the mass $M_{H}$ and angular momentum $J_{H}$, as measured at the event horizon, rather than at spacial infinity. These quantities will play a role in the interpretation of our results below. Following Ref. \cite{Delgado_Herdeiro_Radu:2016}, we can obtain these quantities as functions of the asymptotic mass ($M$) and angular momentum ($J$) for the KNBH. They are given by:
	\begin{widetext}
		\begin{align}
		&M_{HKN}=M\left(1-\frac{Q^{2}}{2 M r_{+}^{KN}}\right)\left[1-\frac{Q^{2}}{a r_{+}^{KN}}\arctan\left(\frac{a}{r_{+}^{KN}}\right)\right],\\
		&J_{HKN}=J \left(1-\frac{Q^2}{2 M r_{+}^{KN}}\right) \left\{1+\frac{Q^2}{2a^{2}}\left[1-\frac{a^{2}+r_{+}^{KN}}{ar_{+}^{KN}}\arctan\left(\frac{a}{r_{+}^{KN}}\right)\right] \right\}.
		\label{JH_MH_KN}
		\end{align}	
	\end{widetext}
	
	\section{Kerr-Sen black hole}
	\hspace{0.5cm} In the low-energy limit, the action for heterotic string theory in the string frame is \cite{Sen1992}:
	\begin{equation}
	\begin{split}
	S = \int d^{4}x\sqrt{-g}e^{-\tilde{\Phi}}\left(R-\frac{1}{8}F_{\mu\nu}F^{\mu\nu}+g^{\mu\nu}\p_{\mu}\tilde{\Phi}\p_{\nu}\tilde{\Phi}\right.\\
	\left. -\frac{1}{12}H_{k\lambda\mu}H^{k\lambda\mu}\right),
	\end{split}
	\end{equation}
	where $\tilde{\Phi}$ is the dilaton field and  $H_{k\lambda\mu}$ is a third-rank tensor field defined as
	\begin{align}
	H_{k\mu\nu}\equiv&\p_{k}B_{\mu\nu}+\p_{\nu}B_{k\mu}+\p_{\mu}B_{\nu k}\nonumber\\
	&-\frac{1}{4}\left(A_{k}F_{\mu\nu}+A_{\nu}F_{k\mu}+A_{\mu}F_{\nu k}\right),
	\end{align}
	with $B_{\mu\nu}$ being a second-rank antisymmetric tensor gauge field.
	
	After applying a solution generating technique to the Kerr solution, Sen obtained a charged rotating BH solution to the equations of motion of the low-energy limit of the heterotic string theory, which became known as the Kerr-Sen (KS) solution.
	Written in Boyer-Lindquist coordinates, in the Einstein frame, the line element of the KS solution is:
	\begin{align}
	ds^{2}=&-\left(1-\frac{2 M r}{\rho^{2}_{KS}}\right)dt^{2}+\rho^{2}_{KS}\left(\frac{dr^{2}}{\Delta_{KS}}+d\theta^{2}\right)\nonumber\\
	&-\frac{4Mra\sin^{2}\theta}{\rho^{2}_{KS}}dtd\phi+\nonumber\\
	&\left[r\left(r+\frac{Q^{2}}{M}\right)+a^{2}+\frac{2 M r a^{2}\sin^{2}\theta}{\rho^{2}_{KS}}\right]\sin^{2}\theta\,d\phi^{2},
	\end{align}
	and the electromagnetic vector potential is given by
	\begin{equation}
	A_{\mu}dx^{\mu}=-\frac{Qr}{\rho^{2}_{KS}}(dt-a\sin^{2}\theta d\phi), 
	\label{A_KS}
	\end{equation}	
	where
	\begin{align}
	&\Delta_{KS}\equiv r\left(r+\frac{Q^{2}}{M}\right)-2 M r+a^{2},\\
	&\rho^{2}_{KS}\equiv r\left(r+\frac{Q^{2}}{M}\right)+a^{2}\cos^{2}\theta.
	\end{align}
	From Eqs.~\eqref{A_KN} and~\eqref{A_KS}, we see that the gauge potential has same form for both KN and KS solutions. However, the function $\rho^2$ is different in both cases. Nonetheless, the leading asymptotic term in $A_t$ is the same. Therefore, an observer at infinity measures $Q$ as the electric charge, for both KNBHs and KSBHs.
	
	The event horizon of the KSBH is located at 
	\begin{equation}
	r^{KS}_{+}=M-Q^2/2 M+\sqrt{\left(M-Q^2/2 M\right)^2-a^2}. 
	\label{rH_KS}
	\end{equation}
	
\begin{figure}[t]
		\centerline{\includegraphics[width=8cm]{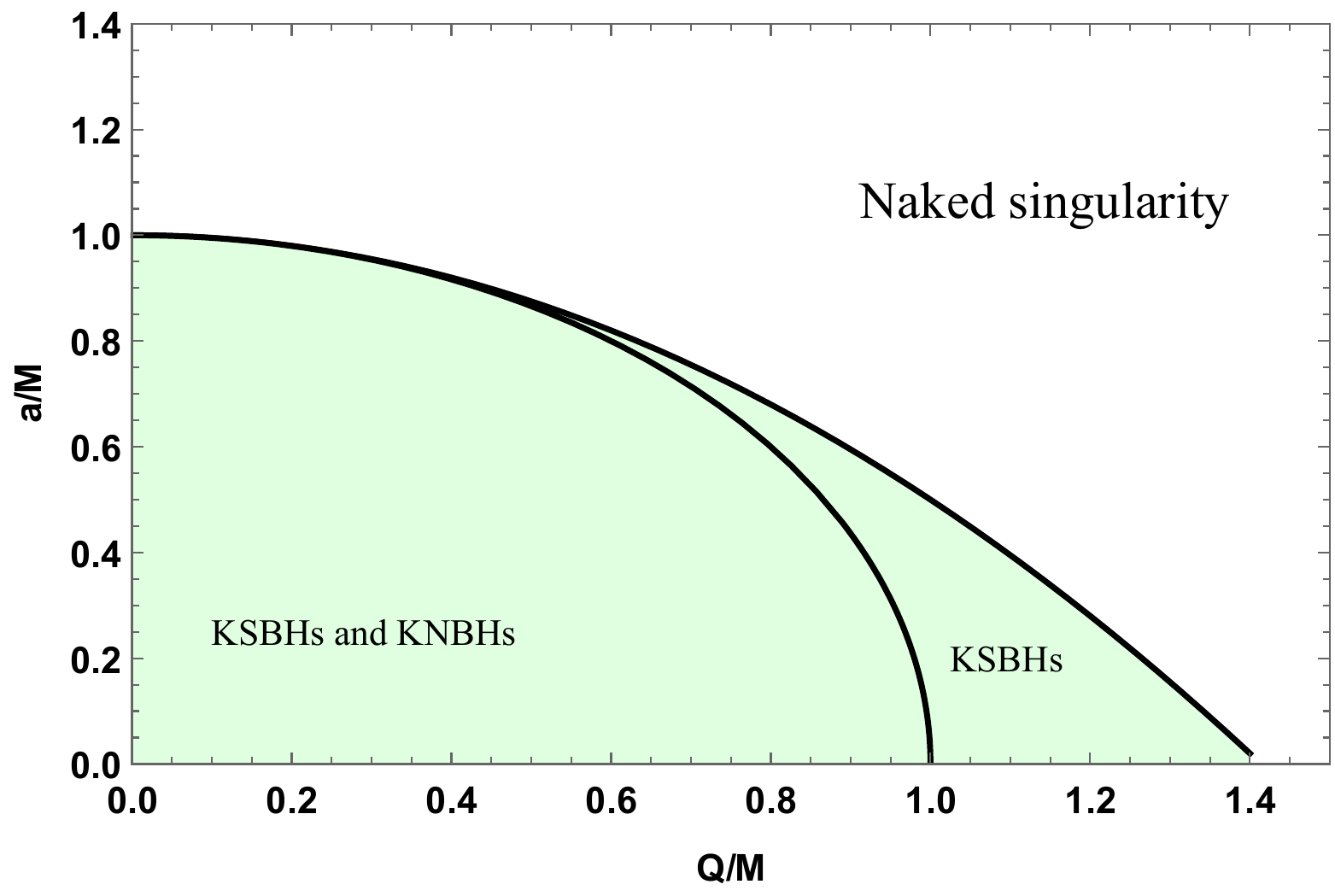}}
		\caption{Dimensionless angular momentum versus charge to mass ratio in KN and KS spacetimes. The solid lines corresponds to extremal BHs~\cite{Hioki:2008}.}
		\label{existenciaKSKN}
	\end{figure}

	In Fig.~\ref{existenciaKSKN} we plot the dimensionless angular momentum, $j\equiv a/M$ versus the charge to mass ratio in KN and KS spacetimes. We see, for instance, that BH solutions are allowed for a wider range of the charge (in units of mass) for the KS case (when compared to the KN case).
	
	
	The KS spacetime is of Petrov type I \cite{Burinskii:1995}, therefore, {\it a priori}, we do not expect to find a Carter-like constant in this geometry. However, surprisingly, the Hamilton- Jacobi equation in KS spacetime is separable for geodesics~\cite{Konoploya:2018}. As given in Ref.~\cite{Lan_Pu:2018}, the null geodesics in the KS spacetime are described by:
	\small
	\begin{align}
	&\rho^{2}_{KS}\dot{r}=\pm\sqrt{R},
	\label{rcarter}\\
	&\rho^{2}_{KS}\dot{\theta}=\pm\sqrt{\Theta},
	\label{thetacarter}\\
	&\rho^{2}_{KS}\dot{t}=\frac{E\left[r\left(r+\frac{Q^{2}}{M}\right)+a^{2}\right]^{2}-2 M r a L}{\Delta_{KS}}-a^{2}E\sin^{2}\theta,
	\label{tcarter}\\
	&\rho^{2}_{KS}\dot{\phi}=-aE +\frac{L}{\sin^{2}\theta}+\frac{a}{\Delta_{KS}}\left[r\left(r+\frac{Q^{2}}{M}\right)+a^{2}E-aL\right],
	\label{phicarter}
	\end{align}
	\normalsize
	where
	\small
	\begin{align}
	&R\equiv\left[aL-E\left(r\left(r+\frac{Q^{2}}{M}\right)+a^{2}\right)\right]^{2}-\Delta_{KS}\left[\left(L-aE\right)^{2}+K\right],\\
	&\Theta\equiv K-\cos^{2}\theta\left[\frac{L^{2}}{\sin^{2}\theta}-a^{2}E^{2}\right].
	\end{align}
	\normalsize
	$E$ and $L$ are, respectively, the total energy and the axial component of the angular momentum of the particle, as measured by an observer at infinity; and  $K$ is a separation constant equivalent to the one originally introduced by Carter \cite{Carterconstant}.
	
	Defining the following impact parameters
	\begin{align}
	&\Phi_{KS} \equiv L/E,\\
	&\eta_{KS} \equiv K/E^{2},
	\end{align} 
	and requiring that $ R (r) = dR (r) / dr = 0 $, the shadow of the KSBH can be determined by the following expressions~\cite{Uniyal_Nandan_Purohit:2018}:
	\small
	\begin{align}
	\Phi_{KS} =&[2 a^2(M^3+M^2 r) +a^2 M Q^2-6 M^3 r^2-2 M^2 Q^2 r+\nonumber\\
	&2 M^2 r^3+3 M Q^2 r^2+Q^4 r]\left[a M \left(2(M^2- M r) -Q^2\right)\right]^{-1},\label{Phi_r}
	\end{align}
	\begin{align}
	\eta_{KS} =&-[r^2 (\left(2 M^2 \left(r^2-Q^2- 3M r\right)+3 M Q^2 r+Q^4\right)^2\nonumber\\
	&-8 a^2 M^4 \left(2 M r+Q^2\right))]\left[a^2 M^2 \left(2 M (r-M)+Q^2\right)^2\right]^{-1}.\label{eta_r}
	\end{align}
	\normalsize
	These parameters define, as for the KNBH, the SPOs, which determine the boundary of the BH shadow. However, we shall not compare the quantities $\Phi$ and $\eta$ in KN and KS spacetimes using the radial Boyer-Lindquist coordinate $r$, since it is not an invariant quantity. For such comparisons, we will use the perimetral radius, defined as~\cite{Cunha_Herdeiro:2018}
	\begin{equation}
	r_{per}\equiv\sqrt{g_{\phi\phi}}|_{\theta=\pi/2}, 
	\label{r_per}
	\end{equation}
	which has a clear geometric meaning.
	%
	%
	
\begin{figure}
		\begin{center}
			\includegraphics[width=5.9cm]{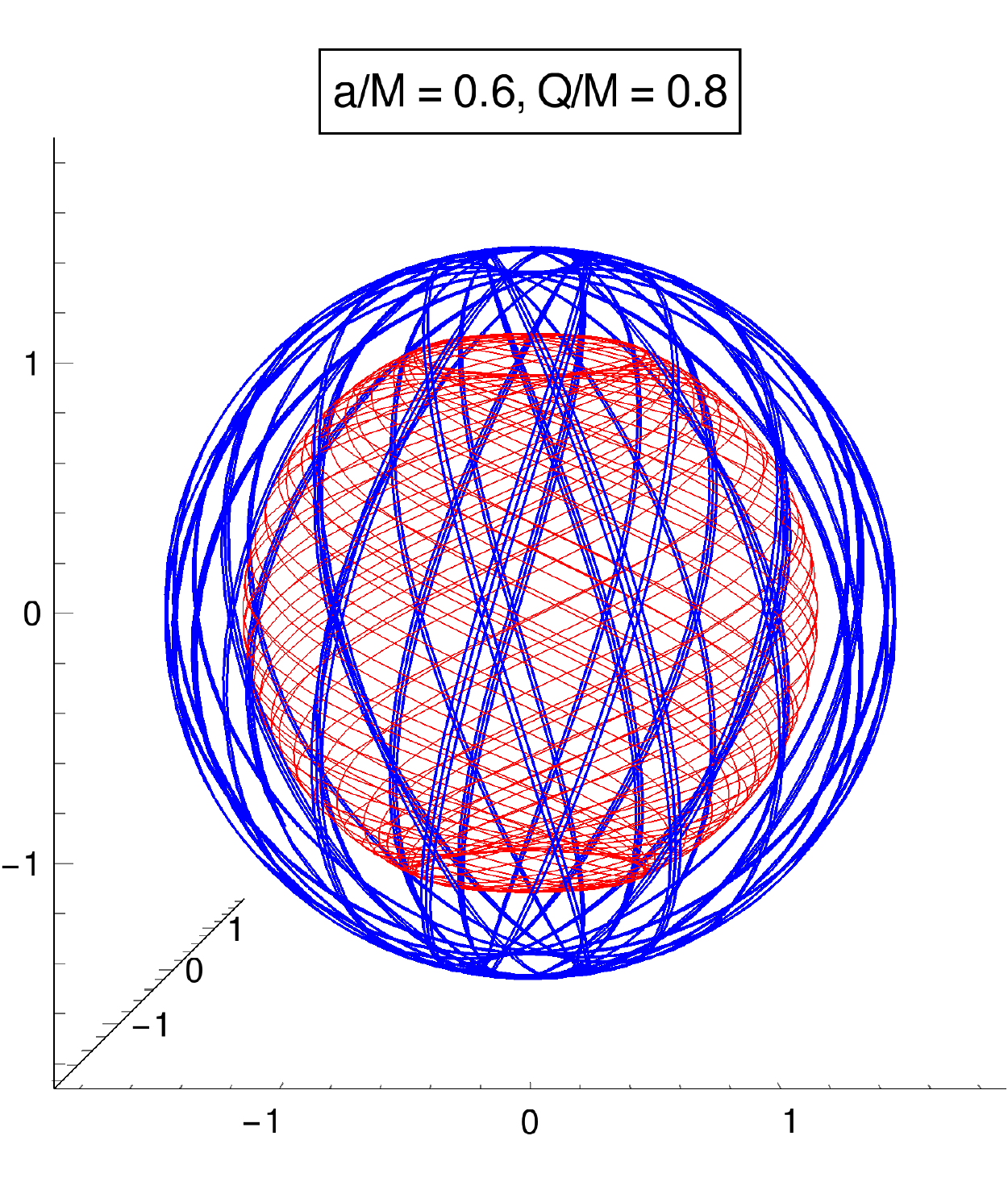}\\
			\includegraphics[width=5.9cm]{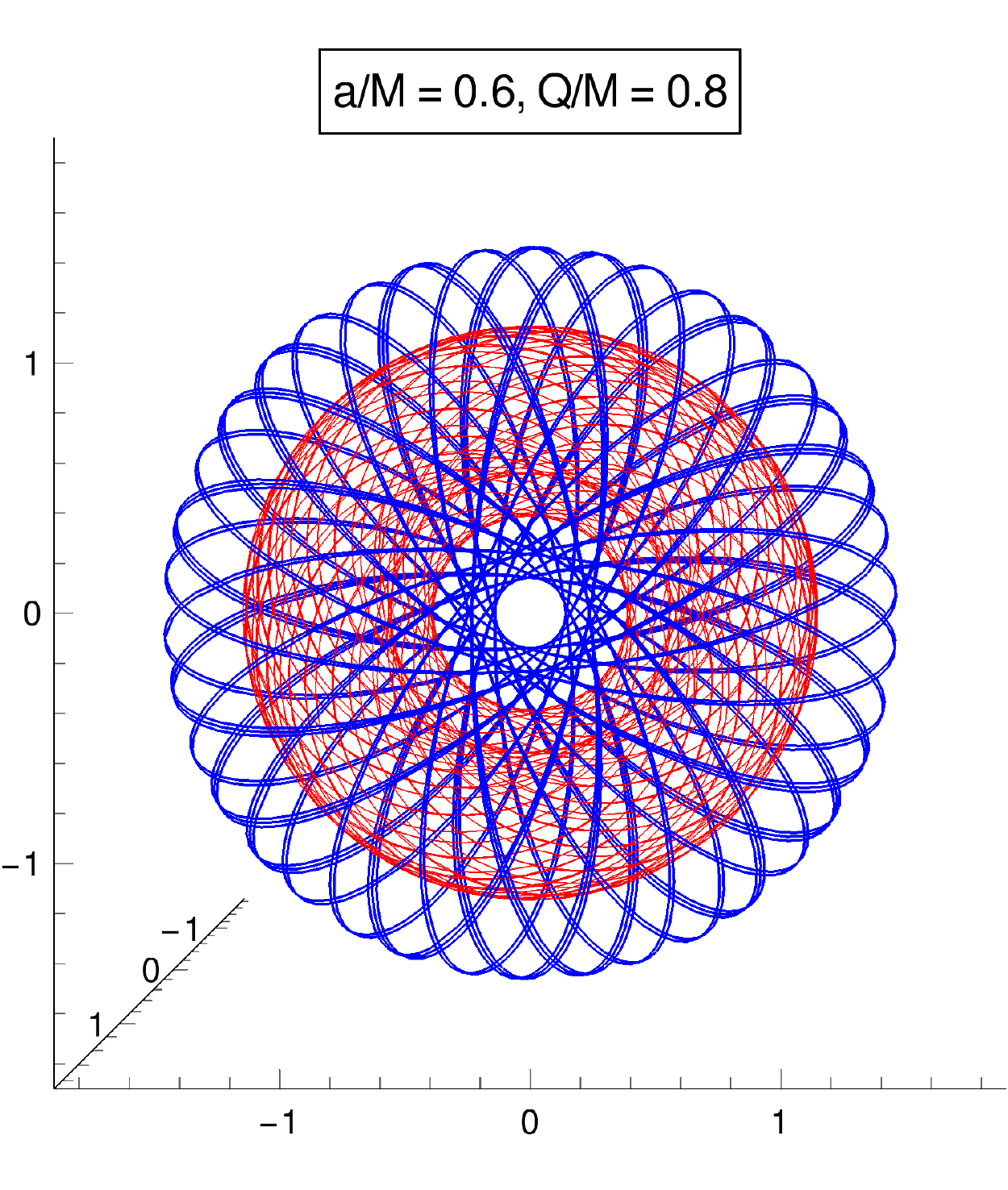}\\
			\includegraphics[width=5.9cm]{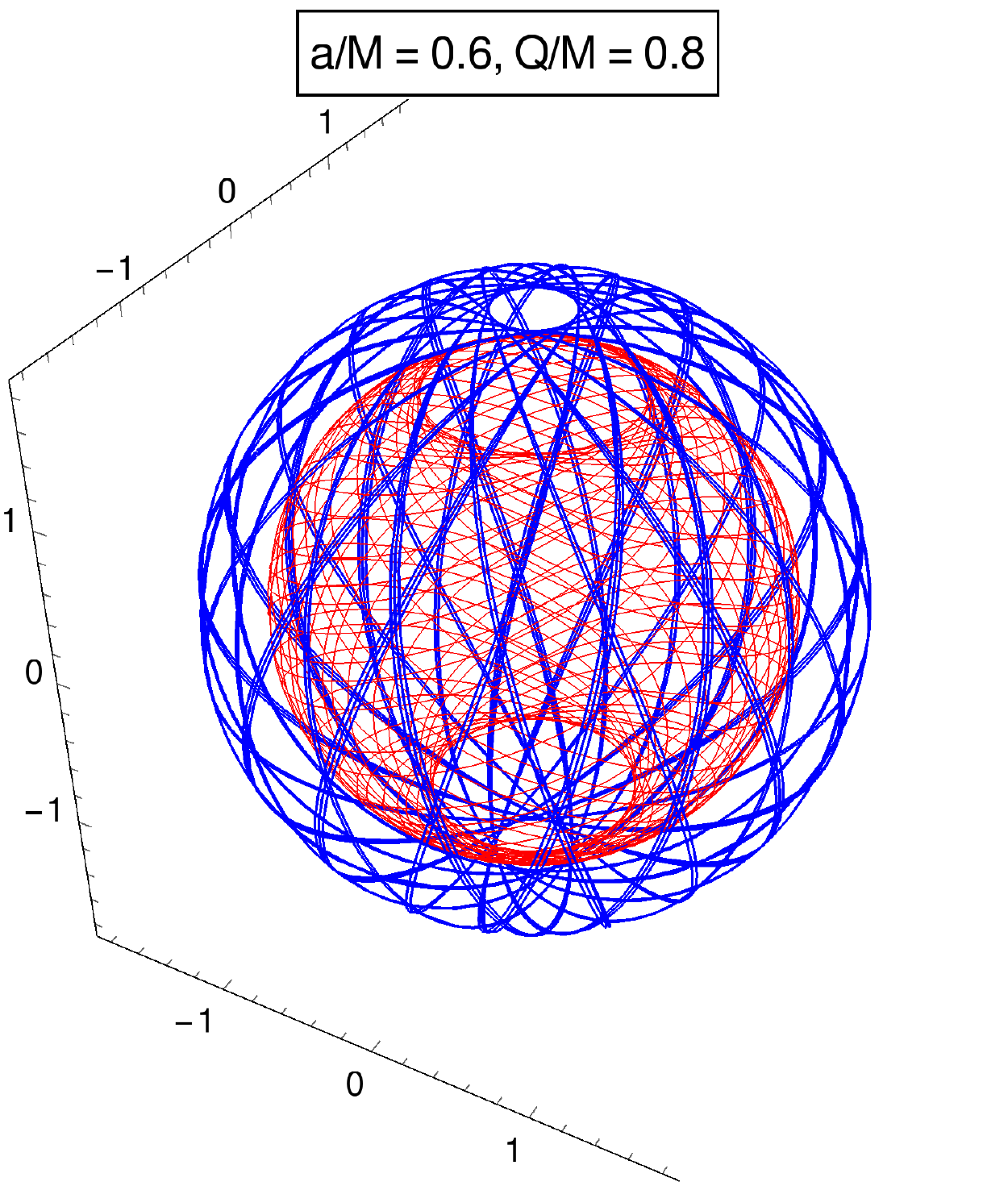}
			\caption{Co-rotating SPOs for KNBH (red lines) and KSBH (blue lines), plotted as functions of the perimetral radius{, for $\Phi/M=2$}. The co-rotating KN SPOs are internal to the KS ones.  
			}
			\label{CorotatingSPO}	
		\end{center}
	\end{figure}

\begin{figure}
		\begin{center}
			\includegraphics[width=6.2cm]{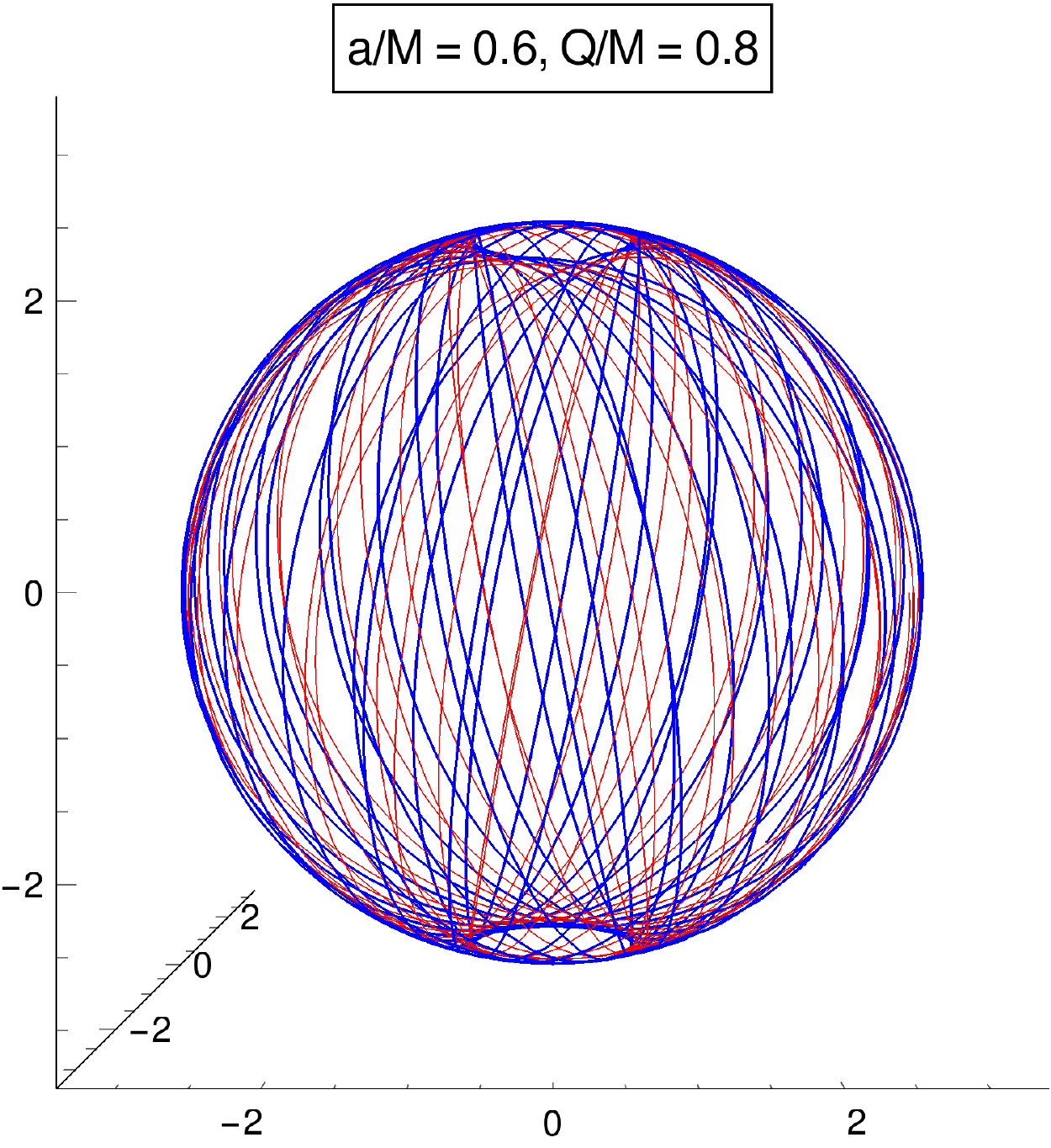}\\
			\includegraphics[width=6.2cm]{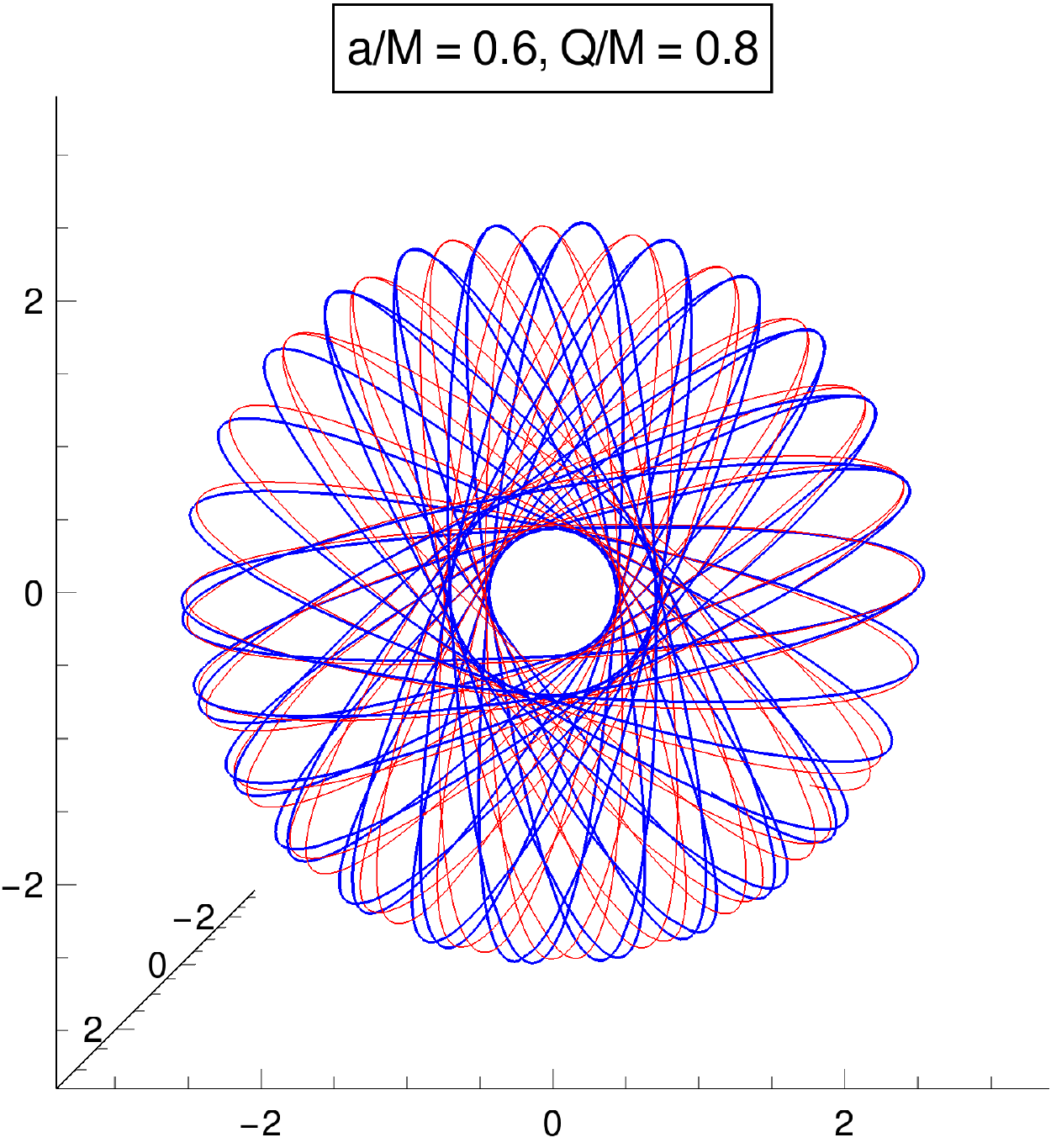}\\
			\includegraphics[width=6.2cm]{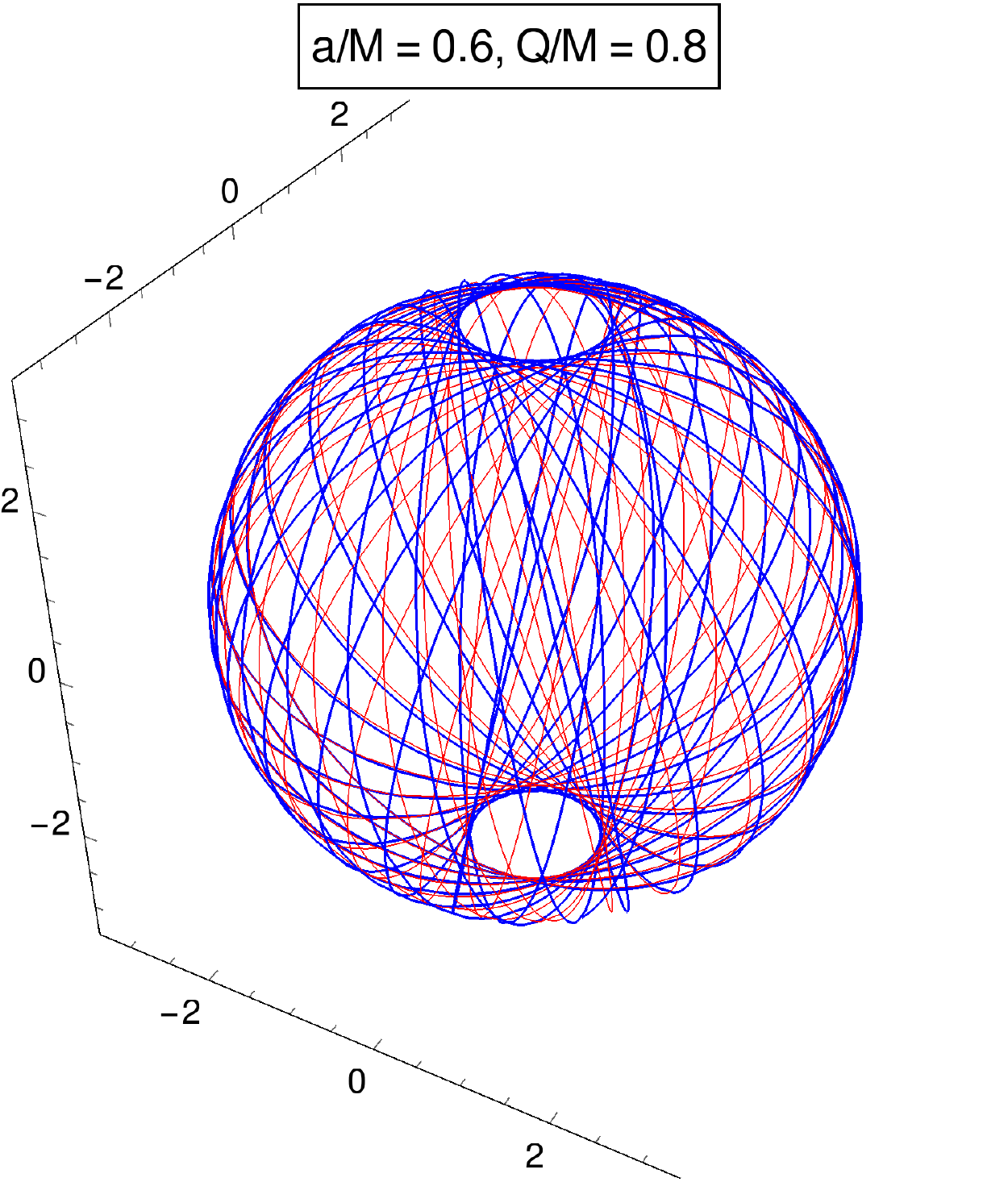}
			\caption{Counter-rotating SPOs for KNBH (red lines) and KSBH (blue lines), plotted as functions of the perimetral radius{, for $\Phi/M=-5$}. As in Fig.~\ref{CorotatingSPO}, the counter-rotating KN SPOs, are internal to the KS ones.}
			\label{CounterrotatingSPO}	
		\end{center}
	\end{figure}

\begin{figure}
		\centerline{\includegraphics[width=8.5cm]{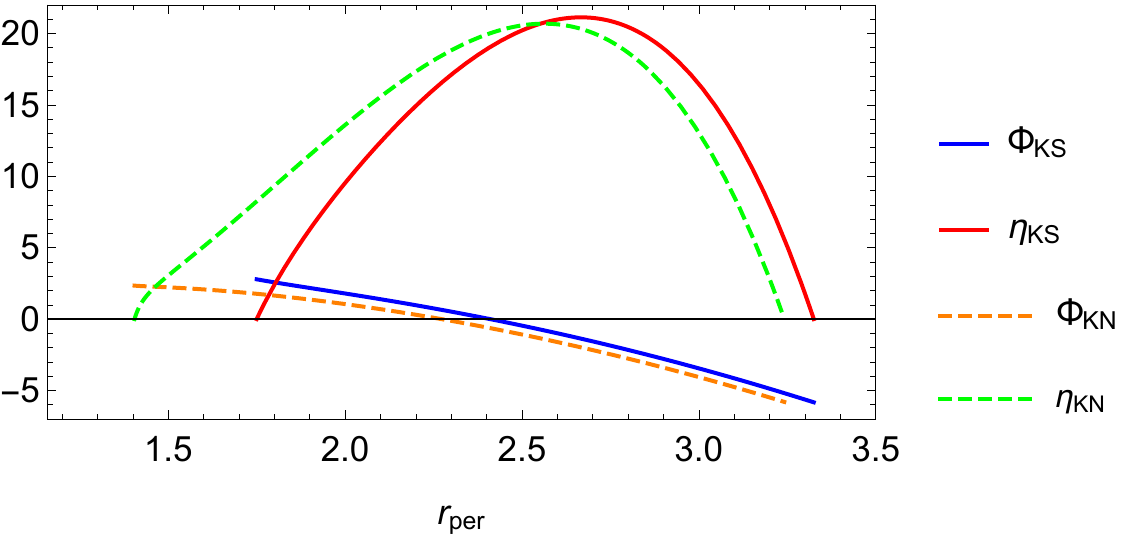}}	
		\caption{Impact parameters of the SPOs, plotted as functions of the perimetral radius, for KSBH (solid lines) and KNBH (dashed lines) spacetimes, with $a/M=0.6$ and $Q/M=0.79$. One can see, in particular, that for light rings, that appear at the extremity of the $\Phi$ curves, the perimetral radius is always larger in the KS case.}\label{phi_eta}
	\end{figure}

	We compare the corresponding SPOs for KNBHs and KSBHs in Figs. \ref{CorotatingSPO} and \ref{CounterrotatingSPO}, plotted as functions of the perimetral radius, for co-rotating and counter-rotating cases, respectively. {In the co-rotating case (Fig. \ref{CorotatingSPO}), the photons have 
		$\Phi/M=2$, whereas in the counter-rotating case (Fig. \ref{CounterrotatingSPO}), 
		$\Phi/M=-5$.} 
	In Fig. \ref{phi_eta} we plot the parameters that determine the SPOs, as functions of the perimetral radius, for $a/M=0.6$ and $Q/M=0.79$. We observe that the co- and counter-rotating orbits in the KS spacetime have a larger radius  than the corresponding ones in the KN spacetime. This already suggests that the shadow of the KSBH will be larger than that of the KNBH for the same global charges, $M,J,Q$.
	
	For the KS spacetime, the mass $M_{H}$ and angular momentum $J_{H}$ as measured at the event horizon are given by \cite{Delgado_Herdeiro_Radu:2016}:
	\begin{widetext}
		\begin{align}
		&M_{HKS}=M\left[\frac{r_{H}^{2}+br_{H}/2}{r_{H}^{2}+br_{H}}-\frac{Q^{2}r_{H}^{2}}{a(r_{H}^{2}+br_{H})^{3/2}}\arctan\left(\frac{a}{\sqrt{r_{H}^{2}+br_{H}}}\right)\right],\\
		&J_{HKS}=J\left[\frac{r_{H}^{2}+3br_{H}/4}{r_{H}^{2}+br_{H}}+\frac{br_{H}}{4a^{2}}-\frac{Q^{2}Mr_{H}^{3}}{a^{3}(r_{H}^{2}+br_{H})^{3/2}}\arctan\left(\frac{a}{\sqrt{r_{H}^{2}+br_{H}}}\right)\right],
		\label{JH_MH_KS}
		\end{align}
	\end{widetext}
	
	where $r_{H}=r_{+}^{KS}$ and $b\equiv Q^{2}/M$.
	\section{Shadows}
	
\begin{figure}[b]
		\centerline{\includegraphics[width=8cm]{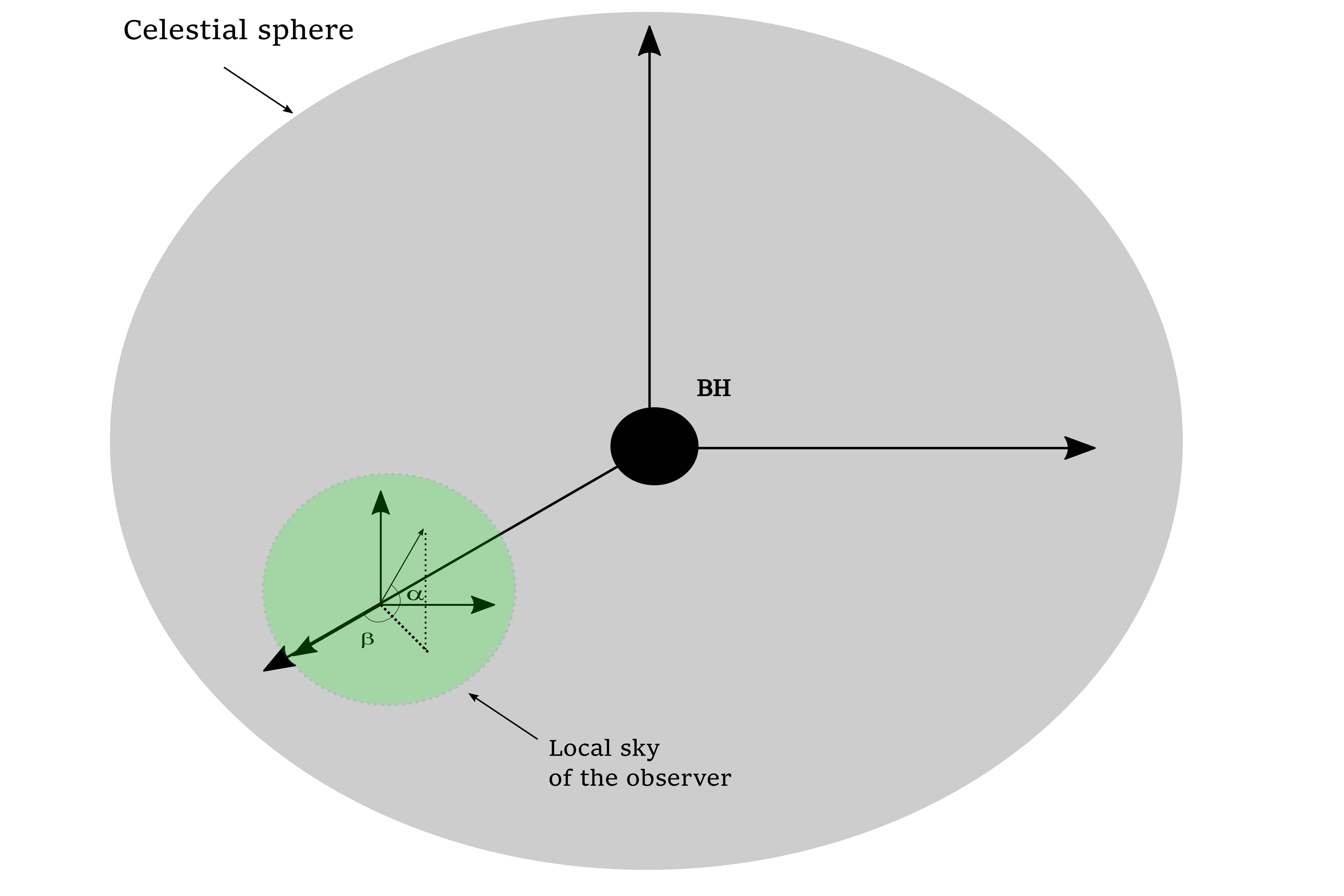}}\caption{Observation angles $\alpha$ and $\beta$ for an observer far away from the BH. The black sphere in the center of the image represents the BH.}
		\label{celestial_coordinate}
	\end{figure}

	\hspace{0.5cm}The shadow of a BH is an observable that can loosely be defined as an absorption cross section of light at high frequencies \cite{Cunha_Herdeiro_Radu}. However, it is observer dependent. In any case, it is the boundary between the captured photon orbits and the scattered photon orbits. For an observer at infinity with latitude coordinate $\theta_{0}$, we can introduce the coordinates $(x,y)$ of the BH shadow edge in the image plane as \cite{Johannsen}:
	\begin{align}
	&x \equiv -\frac{\Phi}{\sin\theta_{0}},\\
	&y \equiv \pm \sqrt{\eta+a^{2}\cos^{2}\theta_{0}-\frac{\Phi^{2}}{\tan^{2}\theta_{0}}}.
	\end{align}
	For an observer at infinity, these coordinates are related with the observation angles ($\alpha, \beta$), that are represented in Fig~\ref{celestial_coordinate}, by the following equations:
	\begin{align}
	&x=-r_{per}\beta,\\
	&y=r_{per}\alpha.
	\end{align}
	They allow us to set up a parametric closed curve in the $x-y$ plane of the observer, representing the contour profile of the shadow of a BH. 
	
	For an observer with $\theta_0=\pi/2$, $\alpha$ is the apparent angular distance away from the axis of symmetry and $\beta$ is the apparent angular distance away from the projected line of the equatorial plane. 
	
\begin{figure}
		\centerline{\includegraphics[width=7cm]{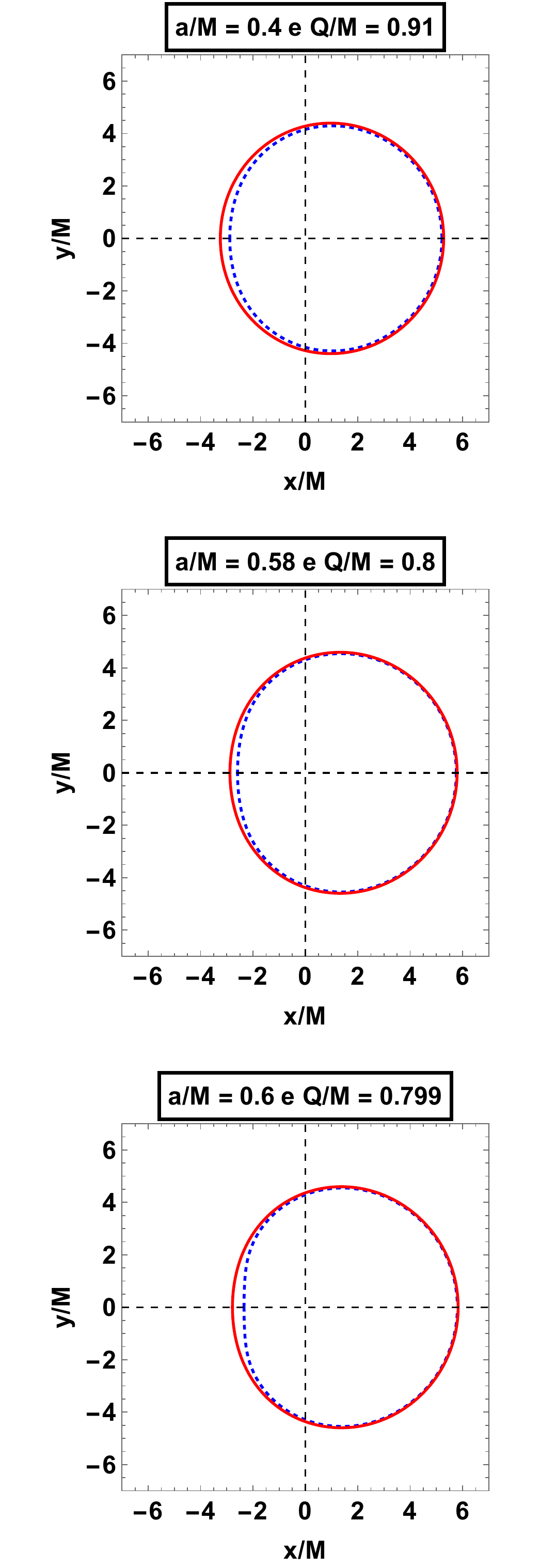}}
		\vspace*{8pt}
		\caption{Shadows of KSBHs (red solid lines) and KNBHs (blue dashed lines), for different values of $a/M$ and $Q/M$. \label{f1}}
	\end{figure}

\begin{figure}
		\centerline{\includegraphics[width=7cm]{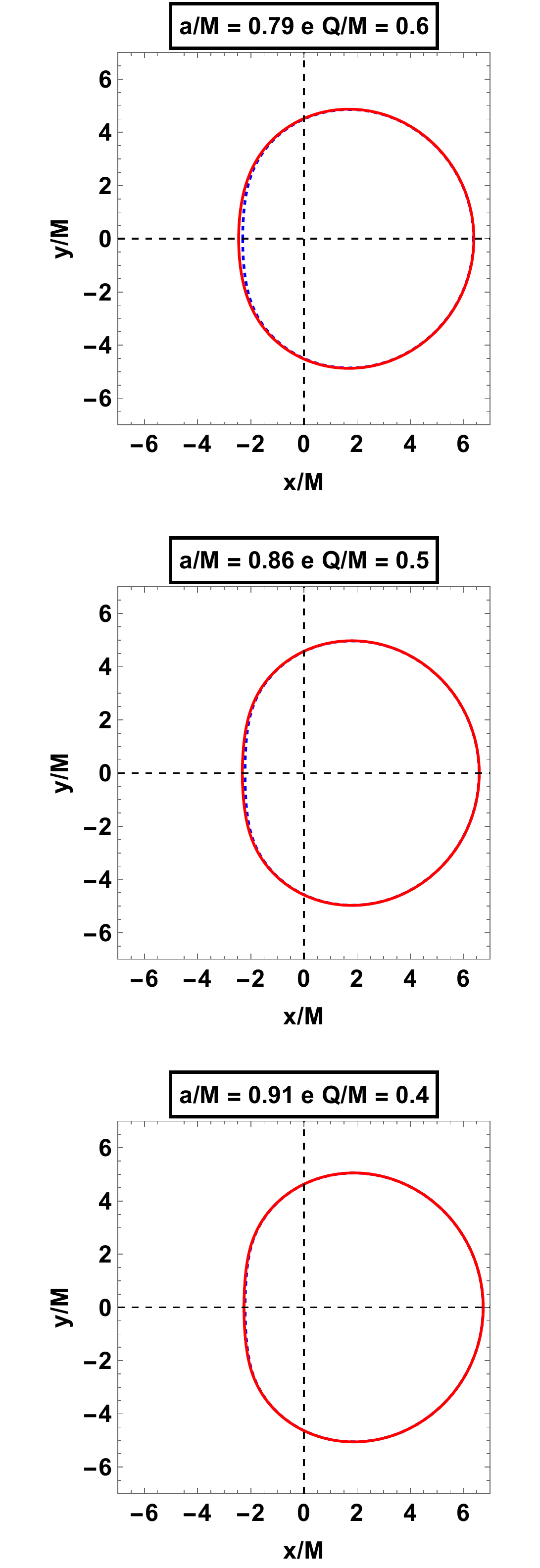}}
		\vspace*{8pt}
		\caption{As in Fig. \ref{f1}, we plot other shadows of KSBHs (red solid lines) and KNBHs (blue dashed lines) for additional choices of $a/M$ and $Q/M$. \label{f2}}
	\end{figure}

	The BH shadow is obtained by plotting $y/M$ versus $x/M$. In Figs. \ref{f1} and \ref{f2} the shadows of KNBHs and KSBHs are represented for different values of $a/M$ and $Q/M$. We notice that the increasing  of the rotation parameter $a$ leads to a loss of symmetry of the BH shadow and enhancing the value of the electric charge leads to a decreasing in the size of the shadows. Besides that, for the same values of the parameters $a$ and $Q$, the shadow of the KSBH is always bigger than the corresponding KNBH one. In both Figs. \ref{f1} and \ref{f2} the observer is localized in the equatorial plane.
	
\begin{figure}
		\centerline{\includegraphics[width=7cm]{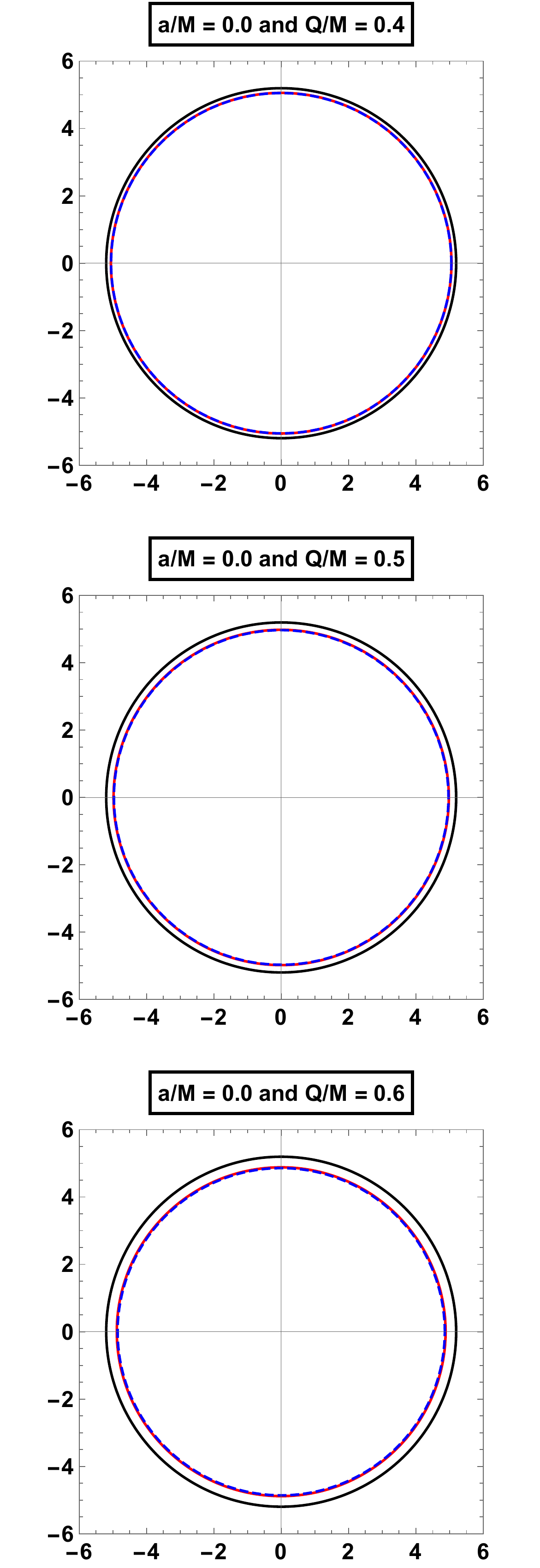}}
		\vspace*{8pt}
		\caption{Shadows of GMBHs (red solid lines) and RNBHs (blue dashed lines), for different values of $Q/M$. The black solid lines represent the shadows of the Schwarzschild BHs. \label{GMRN1}}
	\end{figure}

\begin{figure}
		\centerline{\includegraphics[width=7cm]{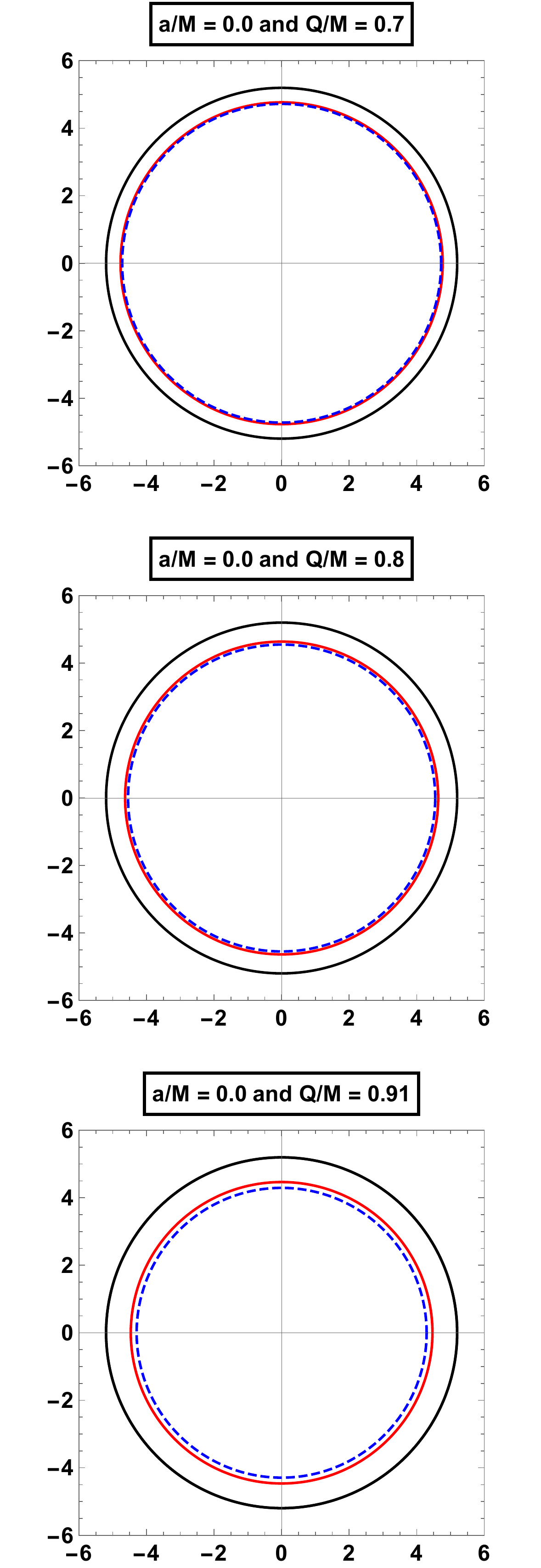}}
		\vspace*{8pt}
		\caption{As in Fig. \ref{GMRN1}, we plot other shadows of GMBHs (red solid lines) and RNBHs (blue dashed lines) for additional choices of $Q/M$. Again, the black solid lines denote the shadows of the Schwarzschild BHs. \label{GMRN2}}
	\end{figure}

	From Figs. \ref{f1} and \ref{f2} we see that as the BH spin decreases,  the relative difference between KSBH and KNBH shadows acquires higher maximum values. If we set $a=0$, the KS solution turns into the Gibbons-Maeda BH (GMBH) \cite{Gibbons_Maeda:1988} whereas the KNBH solution turns into RNBH. In order to understand the charge effect solely, we plot in Figs. \ref{GMRN1} and \ref{GMRN2} the shadows of GMBHs and RNBHs. We can see that, in the static limit, the stringy solution has a larger shadow than the RN solution, although the difference is small.
	
	\section{Shadow radius}
	
	\hspace{0.5cm} Following Ref.~\cite{maeda}, in order to characterize the shadow of a BH, one can define an observable, $R_{s}$, as the radius of the shadow, given by the radius of an approximate circle passing by three points: the top position ($x_{t}, y_{t})$, the bottom position $(x_{b}, y_{b})$, and the point corresponding to the unstable retrograde circular orbit seen from an observer on the equatorial plane $(x_{r}, 0)$. As pointed out in Ref.~\cite{maeda}, one of the advantages of measuring $R_{s}$ is that we can use it to quantify the spin of the BH. The radius $R_{s}$ can be calculated by:
	\begin{equation}
	R_{s}=\frac{(x_{t}-x_{r})^{2}+y^{2}_{t}}{2|x_{t}-x_{r}|}.
	\end{equation}
	This observable can be used as a first step in the development of more complex models.
	
	In order to quantify the contrast between the radius $R_{s}$ of the KSBH and of the KNBH, one can define the relative radius difference as:
	\begin{equation}
	\delta R \equiv \frac{|R_{s}^{KS}-R_{s}^{KN}|}{R_{s}^{KS}},
	\end{equation}
	where $R_{s}^{KS}$ and $R_{s}^{KN}$ are, respectively, the radius of the shadow of the KSBH and of the KNBH.
	
\begin{figure*}
		\centerline{\includegraphics[width=12cm]{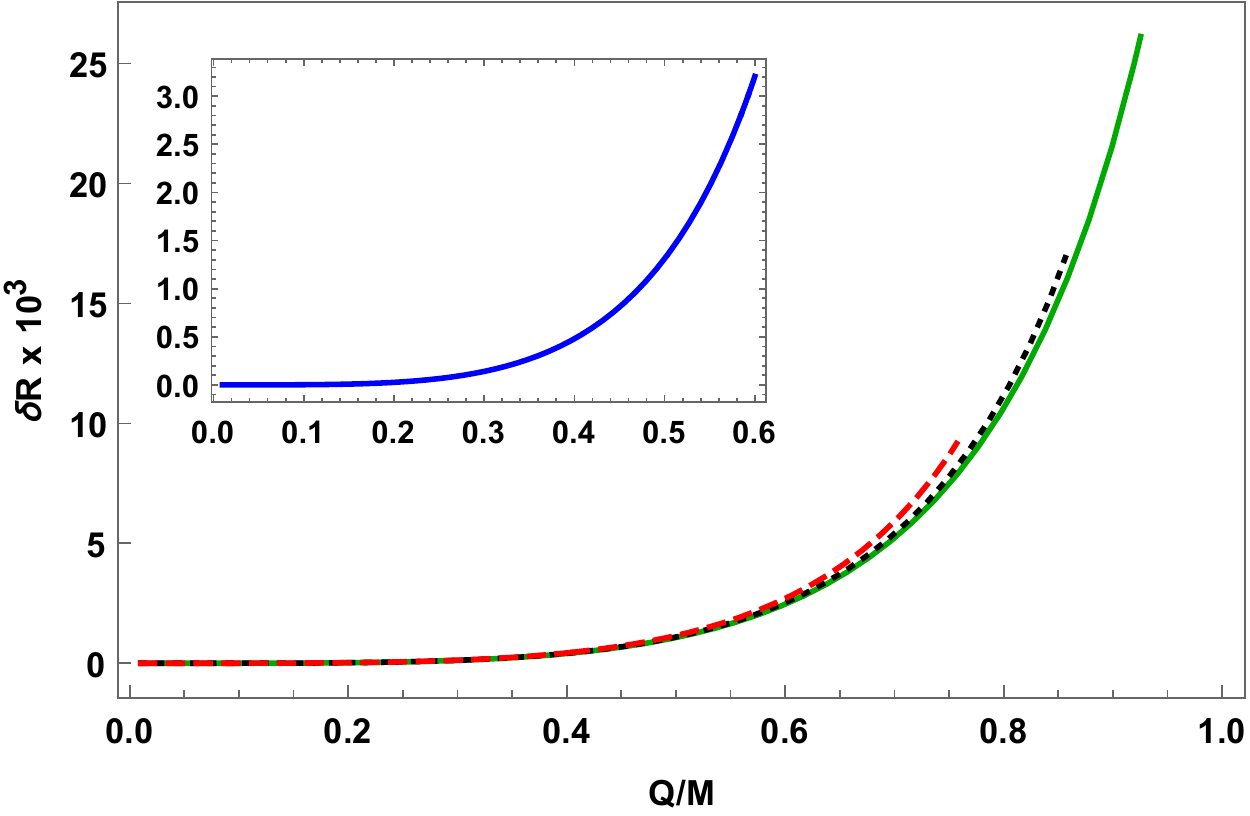}}
		\vspace*{8pt}
		\caption{Relative radius difference, as a function of $Q/M$, for $a=0$ (green solid line), $a/M=0.4$ (black dotted line), $a/M=0.6$ (red dashed line) and  $a/M=0.8$ (blue solid line, in the inset). 
		}
		\label{f3}
	\end{figure*}

	In Fig. \ref{f3} we plot the relative radius difference as a function of the charge $Q$, for different values of $a/M$. We see  that for $a/M=0.4$, the relative difference between the radius of both shadows is very small. The highest difference occurs for the maximum value of the charge for a KNBH, according to Eq.~\eqref{rH_KN}, being around $1.7\%$ from the radius of KNBH. As the spin increases, we note that the contribution of the charge decreases, reaching a deviation of, approximately, $1.2\%$ and $0.32\%$, as it can be seen in the red and blue lines of Fig. \ref{f3}, respectively. This difference is so small that there would be no hope that EHT-like observations could distinguish between the two models of BHs, even if charged BHs were astrophysically relevant. In any case, the conclusion is that the radius is always larger for the stringy BHs, for the same $M,J,Q$. Fig. \ref{f3} also confirms that the largest differences are obtained in the spinless case and for the largest charge to mass ratio.
	
	A tentative explanation for the smaller KNBH shadows comes from comparing the dimensionless angular momentum calculated in the event horizon for the KNBH and the KSBH cases. It is known that the Kerr shadow decreases as the parameter $j\equiv J/M^{2}$ increases. Moreover, it becomes more $D$-shaped and hence less circular, when observed from the equatorial plane.  In the vacuum Kerr spacetime, $M_{H}$ and $J_{H}$ are equal to the asymptotic mass and angular momentum $M$ and $J$, respectively. This is not true for the case of KN and KS spacetimes: $M\neq M_{H}$ and $J\neq J_{H}$ for these electrically charged BHs. Therefore, it is interesting to compare $j_{H}\equiv J_{H}/M_{H}^2$ of KSBHs and KNBHs with the same global charges ($M$, $J$, $Q$).
	
\begin{figure*}
		\centerline{\includegraphics[width=14cm]{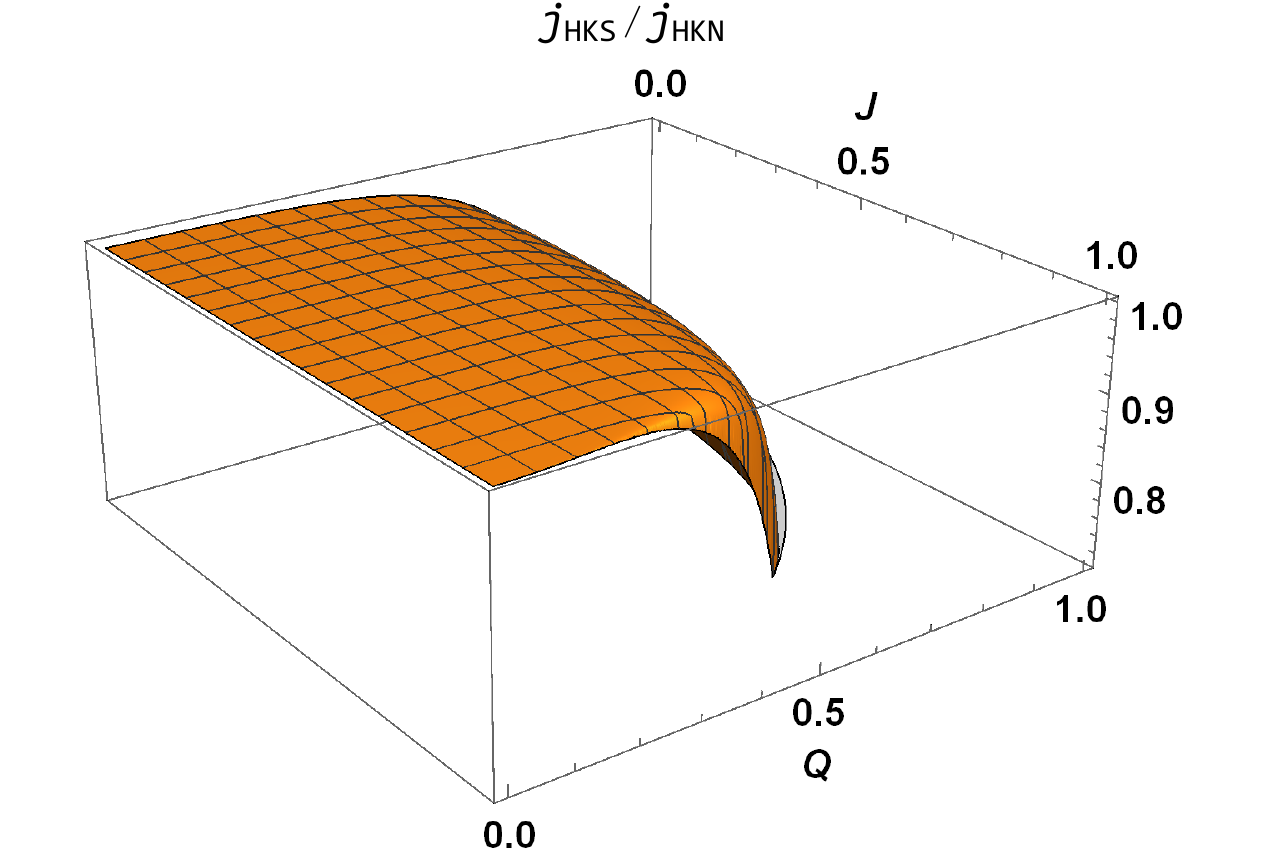}}
		\caption{Ratio of the horizon dimensionless spins, $j_{H}=J_{H}/M_{H}^2$, for KS and KN spacetimes, as a function of the asymptotic dimensionless spin $J$ and charged to mass ratio $Q$, since we have set $M=1$. This ratio is always below unity. Thus, for the same asymptotic global charges $M,J,Q$, the horizon dimensionless spin is always smaller for KSBHs.}
		\label{razao}
	\end{figure*}

	In Fig. \ref{razao} we exhibit $j_H$ for KSBHs and KNBHs, as a function of $J$ and $Q$,  fixing $M=1$. We note that the quantity $j_{HKN}$ is always larger than the corresponding quantity $j_{HKS}$. Thus the stringy BHs store a larger fraction of the dimensionless spin outside the horizon than the KNBH. Since the horizon dimensionless spin of KSBHs is smaller than the one of KNBHs with the same global charges, this justifies why they have a larger shadow, in the same way that for Kerr BHs, the shadow size decreases with increasing dimensionless spin. This interpretation is corroborated, in particular, by (the bottom panel of) Fig.~\ref{f1}, where one observes that the KNBH shadow is not only smaller, but notably more $D$-shaped than the corresponding KSBH case, as one would expect from a horizon which carries a higher dimensionless spin. 
	
	\section{Final Remarks}
	
	\hspace{0.5cm} We have studied SPOs and shadows of KSBHs and KNBHs. Some of our main conclusions are:
	\begin{itemize}
		\item[$\bullet$] From the behavior of the parameters of the SPOs it follows that the co- and counter-rotating SPOs in the KNBH have a smaller perimetral radius than in the KSBH case.
		\item[$\bullet$] The presence of the charge has a clear and similar influence on the apparent size of the shadows in both cases. 
		\item[$\bullet$] The KSBH always has a larger shadow than the KNBH, for the same global, asymptotically measured parameters, $M,J,Q$ and observation conditions. 
		\item[$\bullet$] For the same global, asymptotically measured parameters, $M,J,Q$, the KSBH has a smaller horizon dimensionless spin than a KNBH. This provides an interpretation for the relative smallness of the KNBH shadow as compared to the KSBH shadow. The former, moreover, has typically a higher deformation away from sphericity, what further confirms this interpretation.
		\item[$\bullet$] Computing the relative radius difference between the shadows of both KNBHs and KSBHs, we obtained that this difference increases as the charge $Q$ increases, although remaining small, being around a few percent. 
	\end{itemize}

	A natural extension of our work is to analyze the lensing in both geometries.
	Let us conclude by remarking that, as discussed in the introduction, together with BH imaging, other observations, namely with gravitational waves, are also probing the true nature of astrophysical BH candidates.
	Due to their electric charge, the BHs in this work are not regarded as accurate models to describe astrophysical BHs. Nonetheless, studies of these BHs in relation to this type of phenomenology can be found in $e.g.$ Refs.~\cite{Jai-akson:2017ldo,Siahaan:2019oik}.

	\begin{acknowledgments}
		\hspace{0.5cm}We thank C. L. Benone, G. E. A. Matsas and M. E. Rodrigues for useful suggestions to the final version of this paper. The authors thank Conselho Nacional de Desenvolvimento Cient\'ifico e Tecnol\'ogico (CNPq) and Coordena\c{c}\~ao de Aperfei\c{c}oamento de Pessoal de N\'{\i}vel Superior (Capes) - Finance Code 001, for partial financial support. P. C. is supported by the Max Planck Gesellschaft through the Gravitation and Black Hole Theory  Independent  Research  Group.  This  work  is  supported  by  the Center  for  Research  and  Development  in  Mathematics  and  Applications  (CIDMA)  through  the Portuguese Foundation for Science and Technology (FCT - Funda\c c\~ao para a Ci\^encia e a Tecnologia), references UIDB/04106/2020 and UIDP/04106/2020.  We acknowledge support  from  the  projects  PTDC/FIS-OUT/28407/2017  and  CERN/FIS-PAR/0027/2019.   This work has further been supported by the European Union’s Horizon 2020 research and innovation (RISE) programme H2020-MSCA-RISE-2017 Grant No. FunFiCO-777740.  The authors would like to acknowledge networking support by the COST Action CA16104.
	\end{acknowledgments}
	{}
\end{document}